\documentclass{FLO_v1}%

\usepackage{graphicx}
\usepackage{upgreek}
\usepackage{multicol,multirow}
\usepackage{amsmath,amssymb,amsfonts}
\usepackage{mathrsfs}
\usepackage{amsthm}
\usepackage[figuresright]{rotating}
\usepackage{appendix}
\usepackage[authoryear,round]{natbib}
\usepackage{ifpdf}
\usepackage[T1]{fontenc}
\usepackage{newtxtext}
\usepackage{newtxmath}
\usepackage{textcomp}
\usepackage{xcolor}

\usepackage{epstopdf, epsfig}
\usepackage{color}
\usepackage{tikz}
\usepackage{afterpage}
\usepackage{pifont}
\usepackage{enumerate}

\usepackage{hyperref}
\definecolor{jourcolor}{cmyk}{1,0.57,0.01,0.38}
\hypersetup{
	colorlinks=true,%
	citecolor=jourcolor,%
	filecolor=jourcolor,%
	linkcolor=jourcolor,%
	urlcolor=jourcolor
}

\theoremstyle{definition}

\usepackage{pdflscape}
\usepackage{rotating}
\usepackage[export]{adjustbox}

\usetikzlibrary{shapes}
\usetikzlibrary{plotmarks}

\graphicspath{{./EPS/}}

\renewcommand{\theenumi}{\alph{enumi}}

\newcommand{\cmark}{\ding{51}}%
\newcommand{\xmark}{\ding{55}}%

\definecolor{LightGray}{rgb}{0.75,0.75,0.75}
\definecolor{C39}{rgb}{1,0.78,0.31}
\definecolor{C64}{rgb}{0.98,0.45,0.06}
\definecolor{C133}{rgb}{0.58,0.20,0.39}
\definecolor{CS}{rgb}{0 0 0}

\newcommand{\solidline}[1][black]{\raisebox{2pt}{\tikz{\draw[-,color=#1,solid,line width = 0.5pt](0,0) -- (6mm,0);}}} 

\newcommand{\dashedline}[1][black]{\raisebox{2pt}{\tikz{\draw[-,color=#1,dashed,line width = 0.5pt](0,0) -- (6mm,0);}}}

\makeatletter
\usetikzlibrary{calc}
\newcommand*\circled[1][]{\tikz[baseline=(char.base)]{
		\node[shape=circle,draw=black,fill=white,inner sep=0pt,line width=1pt,minimum height={\f@size*1.1},] (char) {\vphantom{WAH1g}{\textcolor{black}{\footnotesize{\textbf{#1}}}}};}}
\makeatother

\articletype{RESEARCH ARTICLE}

\DOI{xxx}

\Year{2021}

\Vol{}

\Price{}

\art-id{FLOxxxxx}

\citearticle{Wangsawijaya, D. D., Nicolai, C., \& Ganapathisubramani, B.}

\begin{document}
	
	\title[Time-averaged velocity and scalar fields of the flow surrounding a group of cylinders]{Time-averaged velocity and scalar fields of the flow surrounding a group of cylinders}
	
	\author[D. D. Wangsawijaya, C. Nicolai, and B. Ganapathisubramani]{D. D. Wangsawijaya$^{1\ast}$ {{\href{https://orcid.org/0000-0002-7072-4245}{\includegraphics{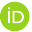}}}}, 
		C. Nicolai$^1$, and B. Ganapathisubramani$^1$}
	
	\address[1]{Aerodynamics and Fluid Mechanics Research Group, Faculty of Engineering and Physical Sciences, University of Southampton SO17 1BJ, United Kingdom}
	
	\corres{*}{Corresponding author. E-mail:
		\emaillink{D.D.Wangsawijaya@soton.ac.uk}}
	
	
	\date{\textbf{Received:} XX 2021; \textbf{Revised:} XX XX 2021; \textbf{Accepted:} XX XX 2021}
	
	\abstract{
		We examine the characteristics of mean flow and scalar concentration characteristics of a turbulent boundary layer flow impinging on a cluster of tall obstacles (which can also be interpreted as a porous obstruction). The cluster is created with a group of cylinders of diameter $d$ and height $h$ arranged in a circular patch of diameter $D$.The solidity of the patch/obstruction is defined by $\phi$ (the total planar area covered by cylinders), which is systematically varied ($0.098 \leq \phi \leq 1$) by increasing the number of cylinders in a patch ($N_c$). A point source is placed at ground level upstream of the patch and its transport around the patch is examined. Time-averaged velocity and scalar fields, obtained from simultaneous planar PIV-PLIF (particle image velocimetry-planar laser-induced fluorescence) measurements, reveal that the characteristics of wake and flow above porous patches are heavily influenced by $\phi$. In particular, we observe that the horizontal and vertical extent of the wake and scalar concentration downstream of the patches decreases and increases with $\phi$, respectively. Here, the recirculation bubble is shifted closer to the trailing edge of the patches as $\phi$ increases, limiting the flow from convecting downstream, decreasing the scalar concentration and virtually `extending' the patch in streamwise direction. As the bubble forms in the trailing edge, vertical bleeding increases and hence the concentration increases above patch where the cylinders appear to be `extend' vertically towards the freestream. 
	}
	
	\maketitle
	
	\begin{boxtext}
		\textbf{\mathversion{bold}Impact Statement}
		
		Canopy flow covers extensive phenomena encountered in both nature and engineering applications. The characteristics of canopies (also referred to as porous obstructions) is known to affect the mixing, momentum, and scalar (e.g. mass and heat) exchange processes in the flow. Examples include flow around urban areas, wind/tidal farms, patches of roughness in ship hulls, in-line heat-exchangers among various others. In almost all these applications, the patch of obstacles/roughness are typically characterised by its geometric properties such as solidity (frontal and plan) or porosity and it is important to understand the flow/scalar characteristics as well as transport processes for varied geometric parameters. For instance, an urban canopy with given distribution of buildings and a pollutant source at ground level could result in different wake/dispersion characteristics depending on the frontal or plan solidity of distribution. This study attempts to carry out simplified experiments that will enable us to understand the intricacies of scalar-momentum transport around such canopies and allow us to build useful models for engineering design and/or prediction for engineering and environmental flows. 
		
	\end{boxtext}
	
\section{Introduction}
\label{sub:intro}

Canopy flow refers to a scenario where a boundary layer develops over and around a  porous surface/obstruction. Examples of such phenomenon are abound in both nature and engineering applications, including but not limited to: flows past aquatic and terrestrial vegetations, atmospheric boundary layers developing over forests, clusters of tall buildings in cities, and wind/tidal farms, which consist of arrays of wind/tidal turbines. These obstructions are known to affect the scalar, momentum, and energy transport process in the flow within the canopy and downstream of the canopies. Aquatic vegetations, for example, provide shielding \citep{kemp2000}, affect the sediment formation and erosion \citep{delangre2008,nepf2012a,nepf2012b,tinoco2016}, and also carbon sequestration in coastal ecosystems \citep{lei2021}. Urban and vegetation canopies induce coherent, multi-scale turbulent motions in the atmospheric boundary layer \citep{katul1997,kanda2006,huang2009a,huang2009b,huang2011,li2011}, which governs flow mixing, momentum, heat and mass exchange within the layer \citep{raupach1981,finnigan2000,poggi2004}, including pollutants and other chemical compounds \citep{poggi2006,aristodemou2018}. While the configuration and tuning of turbines in wind/tidal farms are vital in maximising power output \citep{vennell2011,myers2012}, they may also alter the meteorological pattern and ecosystem of the installation sites. For example, installation of wind turbines has been shown to affect the heat distribution and humidity of the surrounding air \citep{baidyaroy2004,rajewski2016}, while wakes generated by tidal turbines can potentially modify sedimentation process and bacterial population in estuarine waters \citep{ahmadian2012}. Considering its extensive industrial applications and environmental impacts, understanding the transport mechanism of the flow surrounding porous obstructions is, therefore, essential.    

\begin{figure}
	\centering
	\includegraphics[width=11cm, height=5.5cm, keepaspectratio]{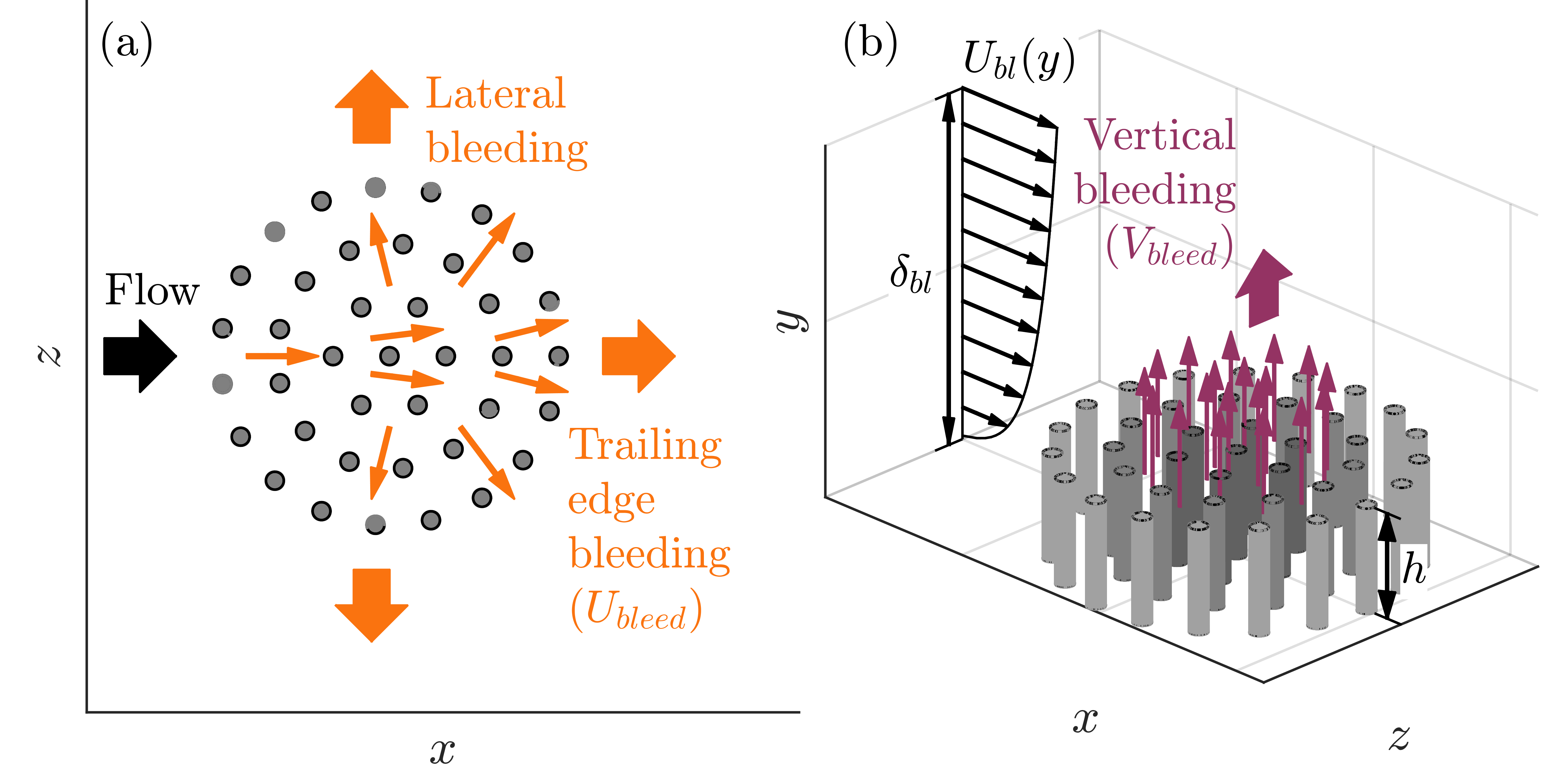}
	
	\caption{Illustrations of (a) lateral and trailing edge bleeding, and (b) vertical bleeding of cylindrical arrays on a circular patch. $U_{bl}$ is the incoming boundary-layer velocity, whose thickness is $\delta_{bl}$, and $h$ is the height of the cylinders.}
	\label{fig:bleed_vel}
\end{figure}

Previous studies on porous obstructions (such as examples given above) typically involved some simplifications from real-life conditions. The obstruction may be simplified as an array of cylinders mounted on an isolated patch with an incoming flow upstream of the cylinders (illustrated in figure~\ref{fig:bleed_vel}). Numerical simulations and experiments have been conducted on the flow surrounding such cylindrical array \citep{ball1996,nepf1997,nepf1999,white2003,tanino2008,nicolle2011,rominger2011,chen2012,zong2012,chang2015,chang2017}, where the cylinders are as tall as, or protruding from, a free surface, as commonly observed in emergent aquatic vegetations. Here, the flow is dominated by K{\'a}rm{\'a}n vortex street in wall-normal axis trailing behind the cylinders and thus can be considered two-dimensional \citep{taddei2016,nicolai2020}. A more general approach is to fully submerge the cylinders in the flow such that $h \ll \delta$ (figure~\ref{fig:bleed_vel}b). Studies have been conducted for cylinders attached to a bottom wall (as in figure~\ref{fig:bleed_vel}b, see also \citealp{chen2013,taddei2016,nicolai2020}) and/or suspended from a free surface \citep{plew2011,tseung2016,zhou2019}. In either cases, the patch is considered as 3-D, as flow mixing occurs on the sides of, downstream of, and above the body. \cite{taddei2016,zhou2019,nicolai2020} have identified three mechanisms in which the flow may escape (`bleed') from a fully submerged porous obstruction: (i) the lateral bleeding from the interior of the patch to the sides (spanwise direction, see figure~\ref{fig:bleed_vel}a), (ii) the trailing edge bleeding, which is responsible in the formation of wake downstream of the body (figure~\ref{fig:bleed_vel}a), and (iii) the vertical bleeding, in which the flow escapes from the interior to the top surface of the patch and towards the freestream (figure~\ref{fig:bleed_vel}b).

It has been reported that the solidity of a porous obstruction is a critical parameter in determining drag \citep{nepf1999,taddei2016,chang2017,zhou2019}, the structure of wake formed by the obstruction \citep{ball1996,nicolle2011,zong2012,chen2012,chang2015,chang2017,zhou2019,nicolai2020}, scalar dispersion \citep{nepf1997,white2003,tanino2008}, and sediment disposition \citep{chen2012,chang2015}. For the cylindrical array shown in figure~\ref{fig:bleed_vel}, the solidity is measured in terms of the planar area covered by cylinders per total surface, $\phi \equiv N_c(d/D)^2$ (also commonly referred to as the solid volume fraction), where $N_c$ is the number of cylinders in a circular patch, $D$ is the diameter of the circular patch, and $d$ is the diameter of the cylinder (see figure~\ref{fig:cases}a,d for graphical descriptions of $D$ and $d$). It has been observed by \cite{nicolle2011,taddei2016,zhou2019} that total drag of the patch $C_D$ increases with the increasing $\phi$ then drops for a solid obstruction ($\phi = 1$, figure~\ref{fig:cases}d). Higher patch solidity also yields shorter and taller wake (in streamwise and wall-normal direction, respectively, see \citealp{nicolai2020}) in the trailing edge, and increases both lateral and vertical bleeding (\citealp{zhou2019}, see also figure~\ref{fig:bleed_vel}). Further, it has been reported in \cite{nicolle2011} that at low $\phi$ ($\phi < 0.05$) each cylinder in an array behaves as a single obstruction, and as the solidity increases to $\phi > 0.15$ the cylindrical array behaves closer to a solid obstruction whose size is equal to that of the patch ($\phi = 1$, figure~\ref{fig:cases}d). Within the range of moderate solidity ($0.05 \leq \phi \leq 0.15$), suppression of vortex street formation downstream of 2-D porous patches has been observed. In its place is a steady wake (constant velocity) region, which is a consequence of flow bleeding. \cite{zong2012} and \cite{chen2012} showed that the streamwise length of this steady wake region scales on $C_D aD$, where $a$ is the frontal area per unit volume $a \equiv N_c d/(0.25\pi D^2)$. Similarly, scaling by $C_D ab$ and $C_D ah$ have also been observed in \cite{rominger2011} for a square patch and \cite{chen2013} for a fully submerged patch, respectively, where $b$ is the half-width of the patch.  

Although the characteristics of the wake and drag of a porous obstruction shown in figure~\ref{fig:bleed_vel} have been discussed in details for both 2-D and 3-D patches, an element that is still missing from these analyses is the intricacy between bleeding and scalar transport in 3-D porous patches. So far, discussions about scalar transport are limited to lateral and longitudinal dispersion in 2-D patches \citep{white2003,tanino2008}, sediment deposition \citep{chen2012,tinoco2016}, qualitative flow visualisation using dye injection \citep{ball1996,rominger2011,zong2012}, and non-simultaneous velocity and scalar measurements due to limitations of the experimental facilities \citep{nepf1999,tanino2008}. Here, a 3-D porous patch serves as a more realistic model of urban landscapes, vegetation canopies, and wind farms, which are typically subsumed within the atmospheric boundary layer. Inclusion of scalar measurements (in addition to the velocity components) has many potential benefits, for example: in air pollution studies, where one might consider cluster of densely populated urban landscape as a porous obstruction and the pollutant as scalar, or in studies about sediment disposition in ocean beds. 

In this study we aim to address the following question: given a point source of scalar located at ground level just upstream  of 3-D porous patches, what are the flow and scalar characteristics in the wake and above the patches for various solidities? To answer this question, we conduct a parametric study of turbulent boundary layers impinging on an array of cylinders mounted on a circular patch (see figure~\ref{fig:bleed_vel}). A point source upstream of the patch is created by releasing a passive fluorescent dye. The solidity $\phi$ is varied by systematically increasing the number of cylinders in the patch, such that $0.098 \leq \phi \leq 1$. This covers both medium and high solidity range described in \cite{nicolle2011}. The axis system $(x, y, z)$ corresponds to the streamwise, wall-normal, and spanwise direction, respectively. Simultaneous planar PIV-PLIF measurements in $x$--$y$ plane provide the instantaneous streamwise--wall-normal velocity components $u$ and $v$, respectively, and scalar concentration field $c$, which can be decomposed further into their time-average and fluctuation about the mean
\begin{equation}
	u(x,y,t) = U(x,y) + u'(x,y,t); \quad v = V + v'; \quad c = C + c'
	\label{eq:decomp}
\end{equation}
It should be noted that due to the choice of the measurement plane, discussion about flow bleeding in the current study is limited only to the trailing edge and vertical bleeding (figure~\ref{fig:bleed_vel}). 

\section{Experimental setup}		
	
\subsection{Test cases}
\label{sub:testcases}

Experiments are carried out in the closed loop water flume at the University of Southampton. The test section has the size of 6250 mm $\times$ 1200 mm (length $\times$ width). In all measurements, the water level is kept constant at $600$ mm and the freestream velocity is set to $U_{\infty} \approx 0.45$ ms$^{-1}$. 

A circular patch with a diameter ($D$) of 100 mm is mounted on the floor of the tunnel at 5500 mm downstream of the test section entrance, which corresponds to the patch Reynolds number $Re_D \equiv U_{\infty}D/\nu \approx 4.5 \times 10^{4}$, where $\nu = 1.0035 \times 10^{-6}$ m$^2$s$^{-1}$ is the kinematic viscosity of water at $20^{\circ}$C. The patch constitutes an array of rigid cylinders, each has the diameter of $d = 5$ mm and height $h = 30$ mm (figures~\ref{fig:bleed_vel}b and ~\ref{fig:cases}a,d). The cylinders are arranged along concentric, evenly-spaced circles whose axis is at the centre of the circular patch, such that the distance between two consecutive cylinders is constant within the patch (see figures~\ref{fig:cases}a--c). The solidity of the patch is parametrically varied by gradually increasing the number of cylinders $N_c$ while keeping $D$ and $d$ constant. Table~\ref{tab:cases} and figures~\ref{fig:cases}(a--c) show three different circular patches: C39, C64, C133, which correspond to $N_c$ = 39, 64, 133, which cover a range of patch density of $0.1 \lesssim \phi \lesssim 0.33$ ($2.5 \lesssim aD \lesssim 8.5$). A solid obstruction case, i.e. a cylinder covering the entire circular patch (such that $N_c = 1$ and $\phi = 1$), is included in addition to the porous cases as case `CS' in table~\ref{tab:cases} and figure~\ref{fig:cases}(d). The solid cylinder in case CS has an equal height $h$ to the cylinder arrays in other test cases.

\subsection{Validation with smooth-wall turbulent boundary layer}
\label{sub:tbl}

\begin{figure}
	\centering
	\includegraphics[width=6.6cm, height=7.2cm, keepaspectratio]{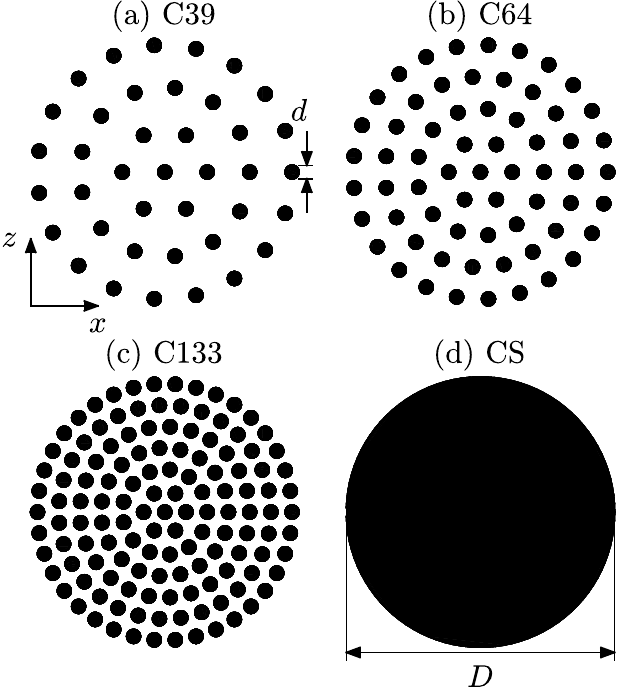}
	\hspace{0cm}
	\includegraphics[width=7cm, height=6.25cm, keepaspectratio]{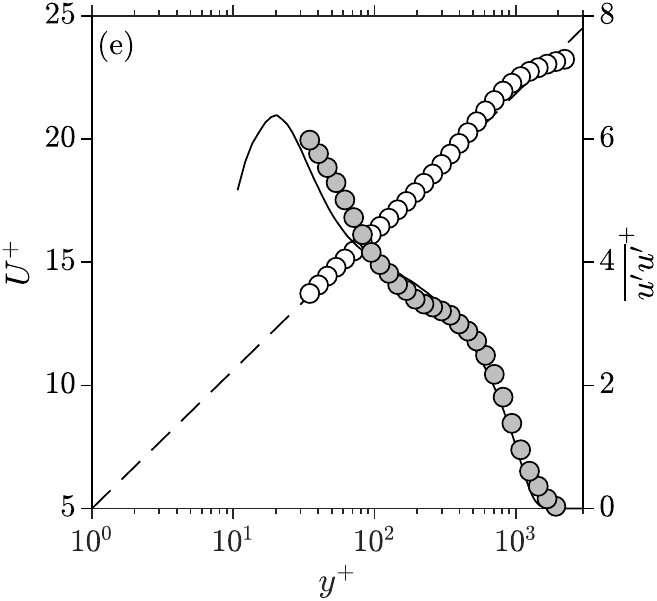}
	
	\caption{Illustration of cylinder array test cases: (a) C39, (b) C64, (c) C133, and (d) the solid case (CS) in $x$--$z$ plane (top view), $d$ is cylinder diameter and $D$ is the circular patch diameter. (e) Mean streamwise velocity $U^+$ ($\circ$) and variance of velocity fluctuation $\overline{u'u'}^+$ (\textcolor{LightGray}{$\bullet$}) of the smooth-wall turbulent boundary-layer (TBL) case as functions of wall-normal location $y^+$ at $Re_{\theta} \approx 3000$, where $\theta$ is the momentum thickness. Dashed line (\dashedline) is $1/(0.41) \log y^+ + 5.0$, solid line (\solidline) is the DNS results of \cite{sillero2014} at $Re_{\theta} = 4000$, filtered according to the spatial resolution of the PIV measurements \citep{lee2016}. Data are downsampled for clarity.}
	\label{fig:cases}
\end{figure}

A smooth-wall turbulent boundary layer (`TBL') case is first tested to validate PIV measurements in the present study. The measurement is conducted in the same test section as other test cases by removing the circular patch. This flow has the Reynolds number (based on momentum thickness $\theta$) of $Re_{\theta} \equiv \theta U_{\infty}/\nu \approx 3000$ and friction Reynolds number $Re_{\tau} \equiv \delta_{bl} U_{\tau bl}/\nu \approx 1500$, where $\delta_{bl} = 82 \pm 3\%$ mm is the boundary-layer thickness, $U_{\tau bl} \equiv \sqrt{\tau_0/\rho} = 0.019 \pm 2\%$ ms$^{-1}$ is the friction velocity for this case, $\tau_0$ is the wall shear stress, and $\rho = 998.12$ kg m$^{-3}$ is the density of water at 20$^{\circ}$C. Figure~\ref{fig:cases}(e) shows the profiles of mean streamwise velocity $U^+ \equiv U/U_{\tau bl}$ and the variance of velocity fluctuations $\overline{u'u'}^+ \equiv \overline{u'u'}/U_{\tau s}^2$ as functions of wall-normal location $y^+ \equiv y U_{\tau s}/\nu$, where subscript `+' denotes viscous scaling and the viscous lengthscale $\nu/U_{\tau s} = 0.053$ mm. The logarithmic region of the mean velocity collapses to the line $1/\kappa \log y^+ + A$, where  $\kappa = 0.41$ is the von K{\'a}rm{\'a}n constant and the $A = 5.0$ is the log-law shift \citep{schlichting2017}. The variance $\overline{u'u'}$ is in good agreement with that obtained from direct numerical simulation (DNS) of \cite{sillero2014} at $Re_{\theta} = 4000$. Note that the DNS results presented in figure~\ref{fig:cases}(e) are filtered according to the spatial resolution of the PIV measurements \citep{lee2016}, $\Updelta x^+ \times \Updelta y^+ = 15 \times 15$ in streamwise and wall-normal directions, respectively. 

\begin{table}
	\centering
	\begin{tabular}{l c c c c c c c c c c c}
		
		\hline
		Cases & $N_c$ & $d/D$ & $h/D$ & $\phi$ & $aD$ & Colour & PIV & PLIF & FOV 1 & FOV 2 & FOV 3\\ 
		\hline
		
		C39 & 39 & 0.05 & 0.3 & 0.098 & 2.483 & \textcolor{C39}{\Large{$\bullet$}} & \cmark & \cmark & \cmark & \cmark & \cmark\\
		C64 & 64 & 0.05 & 0.3 & 0.160 & 4.074 & \textcolor{C64}{\Large{$\bullet$}} & \cmark & \cmark & \cmark & \cmark & \cmark\\
		C133 & 133 & 0.05 & 0.3 & 0.333 & 8.467 & \textcolor{C133}{\Large{$\bullet$}} & \cmark & \cmark & \cmark & \cmark & \cmark\\
		CS & 1 & 1 & 0.3 & 1 & 1.273 & \textcolor{CS}{\Large{$\bullet$}} & \cmark & \cmark & \cmark & \cmark & \xmark\\
		TBL & -- & -- & -- & -- & -- & -- & \cmark & \xmark & \cmark & \cmark & \xmark\\
		\hline
	\end{tabular}
	\caption{List of cylinder array test cases. $N_c$ is the number of cylinders in a patch, $d$ and $D$ are the diameters of the cylinders and the patch, respectively, $h$ is the height of the cylinders, $\phi$ is patch solidity, and $a$ is the frontal area per unit volume. `PIV' and `PLIF' columns show the availability of PIV and PLIF images of each test case. The last three columns show the availability of those images in each FOV.}
	\label{tab:cases}
\end{table}

\subsection{Particle image velocimetry (PIV) and planar laser-induced fluorescence (PLIF)}
\label{sub:exp}

\begin{figure}
	\centering
	\includegraphics[width=14.4cm, height=7.4cm, keepaspectratio]{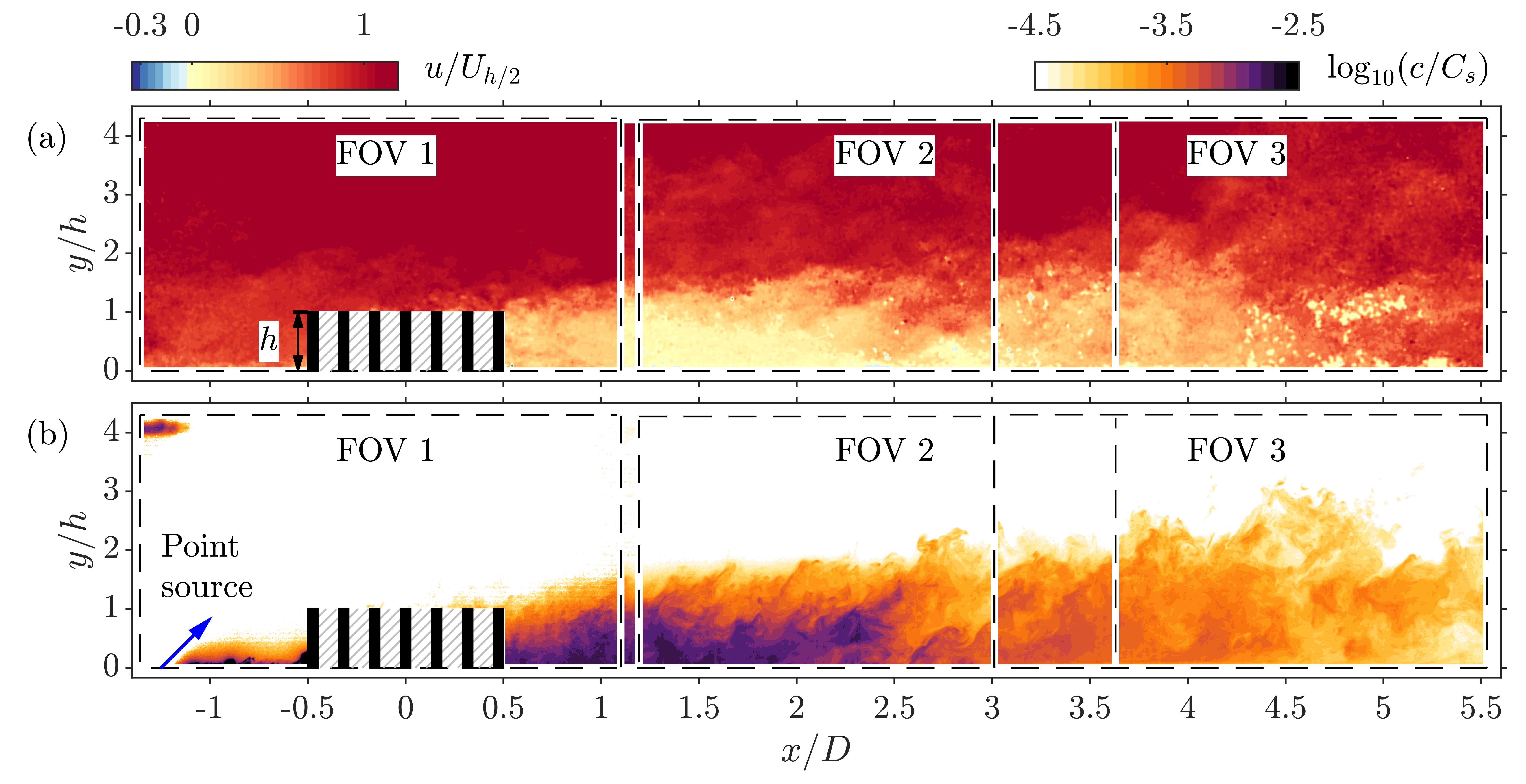}
	\caption{Contours of instantaneous (a) streamwise velocity $u/U_{h/2}$ and (b) scalar concentration $c/C_s$ (in logarithmic scale) of case C39, obtained from (a) PIV and (b) PLIF measurements, respectively, where $U_{h/2}$ is the incoming velocity at half-height of the cylinders. Dashed lines show the extent of FOV 1, 2 and 3. Black rectangles illustrate the cylinder array of height $h$ mounted on a circular patch. The origin $x = 0$ is at the centre of the patch and $y = 0$ is on the floor of the tunnel. Measurement data are not available in the hatched regions (between the cylinders). In (b), blue arrow shows the location of the scalar point source.}
	\label{fig:setup}
\end{figure}

For simultaneous planar PIV and PLIF measurements, the measurement domain is located in the streamwise--wall-normal ($x$--$y$) plane, slicing the centerline of the circular patch (figure~\ref{fig:setup}) and illuminated by Nd:YAG 100 mJ pulsed laser. The measurement domain is created by combining 3 successive experiments in field-of-view (FOV) 1, FOV 2, and FOV 3, which extends downstream of the circular patch. Dashed lines in figure~\ref{fig:setup} illustrate these FOVs and the overlaps between each FOV. It should be noted that although the PIV and PLIF measurements are conducted simultaneously in each FOV, these measurements are not conducted simultanenously between FOV 1, 2 and 3. PIV and PLIF images are recorded first in FOV 1, then the cameras and optics are traversed downstream to form FOV 2, then FOV 3. Measurements are conducted in all three FOVs for the cylindrical array cases (C39, C64, and C133), and only the first two FOVs for the baseline cases (CS and TBL, see table~\ref{tab:cases}). 

In each FOV, PIV images are recorded by a pair of 4 MP CMOS cameras, equipped with wavelength filters to filter out the PLIF signal. As many as 3000 image pairs are acquired for each case at the rate of 4 Hz, which corresponds to the boundary-layer turnover rate of $TU_{\infty}/\delta_{bl} = 1.4$. The timing between a pair of PIV images is $\sim$1 ms ($t^+ \equiv \Updelta t U_{\tau bl}^2/\nu \approx 0.32$). Image cross-correlation and post-processing of the resulting velocity vectors are conducted using a commercial software (DaVis, LaVision GmbH), with the final window size of $24 \times 24$ pixels (50\% overlap) and image pixel size of 15 pixels/mm, which corresponds to the spatial resolution of $\Updelta x^+ \times \Updelta y^+ = 15 \times 15$ for case TBL. Each FOV has the size of 253 mm $\times 126$ mm ($x \times y$), which approximately extends 2.5 times the patch diameter (2.5$D$) in streamwise and 4$h$ in height. When the three FOVs are stitched together, the total length in $x$ is 689 mm ($6.89D$). 

For PLIF measurements, a point source is created upstream of the circular patch by injecting Rhodamine 6G dye solution (Schmidt number $Sc \approx 2500$) to the flow at the rate of 30 cc/min. The dye is supplied by gravity through a thin tube (3 mm diameter) embedded at the floor of the tunnel at 45$^{\circ}$ angle. The tube is located at 125 mm upstream from the centre the patch ($x/D = 1.25$, see figure~\ref{fig:setup}b). PLIF images are acquired using a 5.5 MP 16-bit sCMOS camera equipped with a wavelength filter to block all light except the PLIF signal \citep{vanderwel2014}. To reduce the uncertainty of the scalar concentration measurements, the laser power is recorded at each pulse by an in-line power energy monitor. To maximise the signal-to-noise ratio of the scalar concentration, the concentration of the dye solution at point source $C_s$ is varied based on the location of the FOV relative to the point source. For FOV 1 (closest to the point source), $C_s = 0.3$ mg/L, while for FOV 2 and FOV 3, $C_s = 10$ mg/L. The measured scalar concentration presented in this study is normalised by $C_s$.

\section{Results and discussions}

\begin{figure}
	\centering
	\includegraphics[width=14.4cm, height=9.2cm, keepaspectratio]{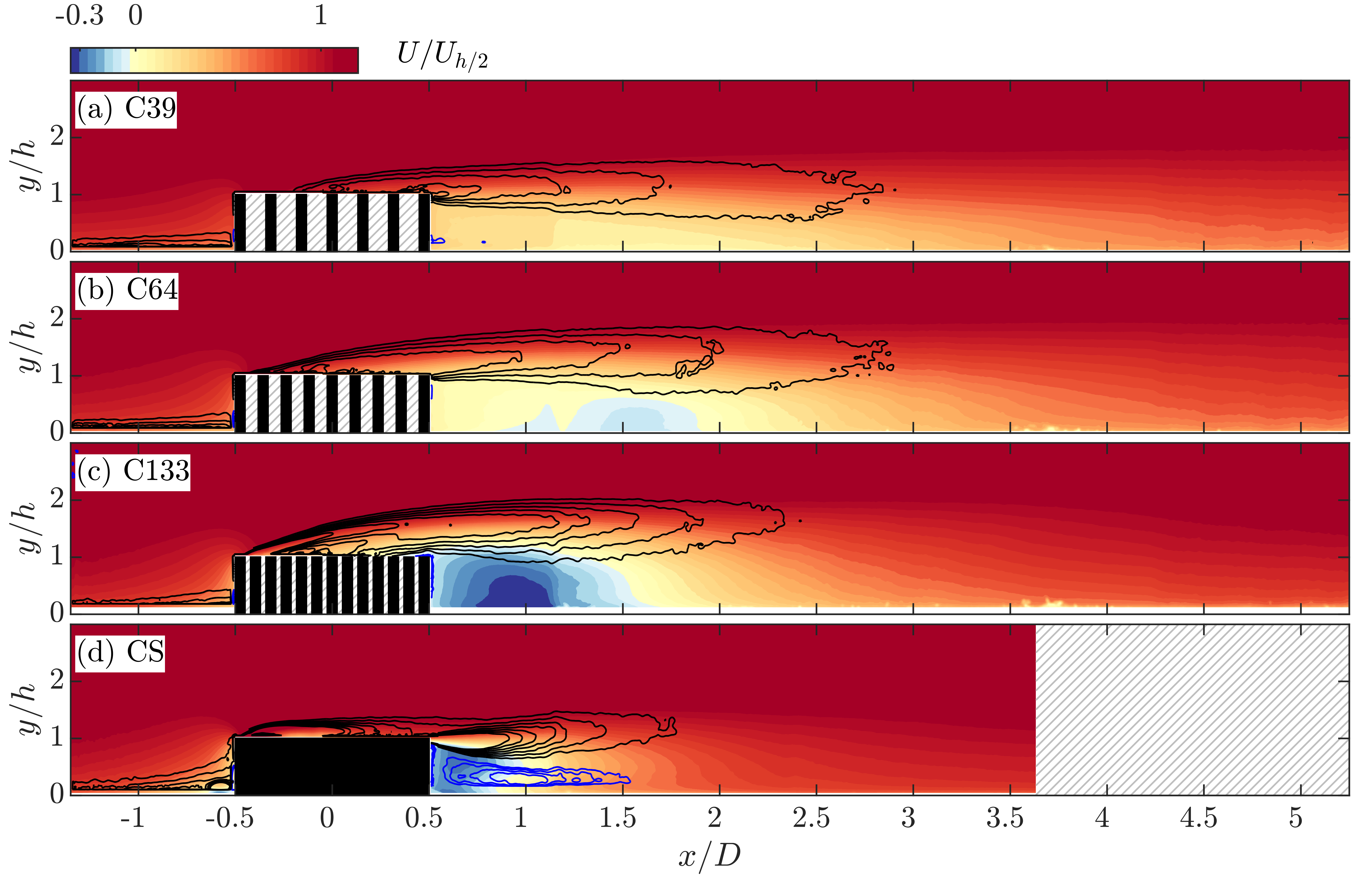}
	\caption{Contours of time-averaged streamwise velocity $U/U_{\infty}$ of case: (a) C39, (b) C64, (c) C133, and (d) CS. Lines are the time-averaged spanwise vorticity $\Omega_y D/U_{h/2}$ with the following contour levels: $-8, -7, ..., -2$ (\solidline) and $0.4, 0.8, ..., 1.6$ (\solidline[blue]).}
	\label{fig:U}
	
	\vspace{0cm}
	\includegraphics[width=14.4cm, height=9.2cm, keepaspectratio]{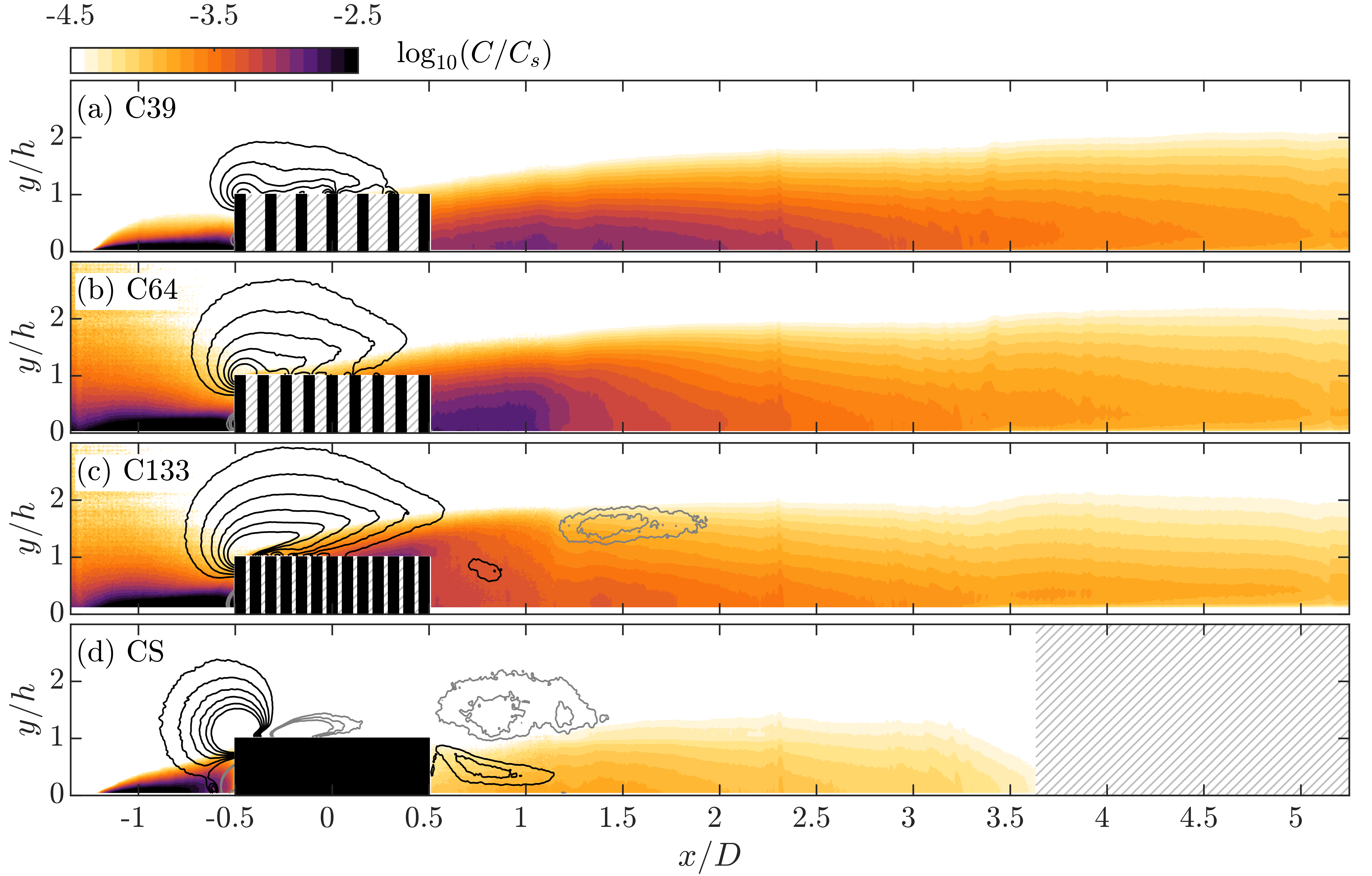}
	\caption{Contours of time-averaged scalar concentration $C/C_s$ (in logarithmic scale) of case: (a) C39, (b) C64, (c) C133, and (d) CS. Lines are the time-averaged wall-normal velocity $V/U_{h/2}$ with the following contour levels: $-0.07, -0.06, -0.05$ (\solidline[LightGray]) and $0.075, 0.1, ..., 0.2$ (\solidline). Measurement data are not available in the hatched regions.}
	\label{fig:C}	
\end{figure}

\subsection{Trailing edge bleeding, the extent of the wake and scalar concentration}
\label{sub:te_bleed}

Figures~\ref{fig:U}(a--d) show the contours of time-averaged streamwise velocity $U$ for all porous test cases and the solid case (CS) listed in table~\ref{tab:cases}. The contours are normalised by $U_{h/2}$, which is defined as the incoming, undisturbed streamwise velocity upstream of the patch ($x/D = -1.3$) at the half-height of the cylinder, $y/h = 0.5$. The figures show the formation of wake for all porous and solid cases in the form of velocity deficit downstream of the patches. The magnitude of the deficit is significant, which signals the presence of patch-sized flow phenomena \citep{taddei2016} as opposed to those of individual cylinders observed in patches with low solidity \citep{chen2012,chang2015,taddei2016,nicolai2020}. In figures~\ref{fig:U}(b--c), blue-shaded contours show $U < 0$, which correspond to the recirculation of the flow downstream of the patch. The strength of the recirculating flow appears to be affected by the solidity of the patch: recirculation is not apparent in case C39 (case with the lowest solidity, figure~\ref{fig:U}a), a weak recirculation then appears in case C64 (figure~\ref{fig:U}b), which grows stronger in case C133 (figure~\ref{fig:U}c). It is also noted that the recirculation bubbles are shifted downstream as solidity decreases. In case C64 (figure~\ref{fig:U}b) the bubble is observed at $1 \lesssim x/D \lesssim 2$, while in case C133 (figure~\ref{fig:U}c) and the solid case CS (figure~\ref{fig:U}d), the bubbles form immediately at the trailing edge of the patches ($x/D = 0.5$). Similar phenomenon has been observed in 2-D patches \citep{chang2015}, 3-D patches \citep{taddei2016,zhou2019}, and porous plates \citep{castro1971}. Contours of positive (upward) wall-normal velocity $V$ above the patches are shown in solid lines in figure~\ref{fig:C}. For the porous cases in figures~\ref{fig:C}(a--c), vertical bleeding is apparent above the patches, characterised by the contours of $V > 0$ at $-0.5 \leq x/D \leq 0.5$ and $y/h \geq 1$. This will be left for further discussion in \S\ref{sub:v_bleed}.  

Superimposed with the contours of $U$ in figures~\ref{fig:U}(a--d) are the contours of time-averaged spanwise vorticity $\Omega_y \equiv \partial V/\partial x - \partial U/\partial y$ normalised by $U_{h/2}/D$. Black solid lines are the contours of $\Omega_y < 0$ (clockwise) and the blue lines $\Omega_y > 0$ (counter-clockwise). Negative $\Omega_y$ show the development of shear layers forming above and downstream of the patches, which appear to be stronger (higher in magnitude) as solidity increases. Similar to the observation in \cite{taddei2016}, here we also observe the elevation of the shear layer further upward from the patches as solidity increases (compare, for example, figure~\ref{fig:U}c to figures~\ref{fig:U}a,b). The lift-off is attributed to vertical bleeding from the interior of the patches, which is noticeably absent in the solid case CS (figure~\ref{fig:U}d) where there is no vertical bleeding. 

Figures~\ref{fig:C}(a--d) show the contours of time-averaged scalar concentration $C$ (normalised by the concentration at the source $C_s$ and presented in logarithmic scale) for all porous test cases and the solid case (CS) listed in table~\ref{tab:cases}. Dark-shaded contours in figures~\ref{fig:C}(a--c) illustrate how the scalar permeates through and downstream of the patches. With less obstructions to block the flow, high scalar concentration (purple-shaded) is observed downstream of the C39 and C64 patch in figures~\ref{fig:C}(a,b), respectively. As blockage ($\phi$) increases, TE bleeding decreases and consequently, the downstream concentration decays (see C133 patch, figure~\ref{fig:C}c). Instead, the scalar escapes the patch by means of vertical bleeding, as shown by higher scalar concentration above the patch ($y/h \geq 1$) in figure~\ref{fig:C}(c). As $\phi$ reaches 1 (solid case), TE and vertical bleeding are non-existent, as evidenced by $C \approx 0$ above the patch and low magnitude of $C$ downstream of the patch (figure~\ref{fig:C}d). Here, the observed concentration is mainly the result of flow mixing in the wake instead of the scalar permeating through the patch.    

\begin{figure}
	\centering
	\includegraphics[width=14.4cm, height=4.6cm, keepaspectratio]{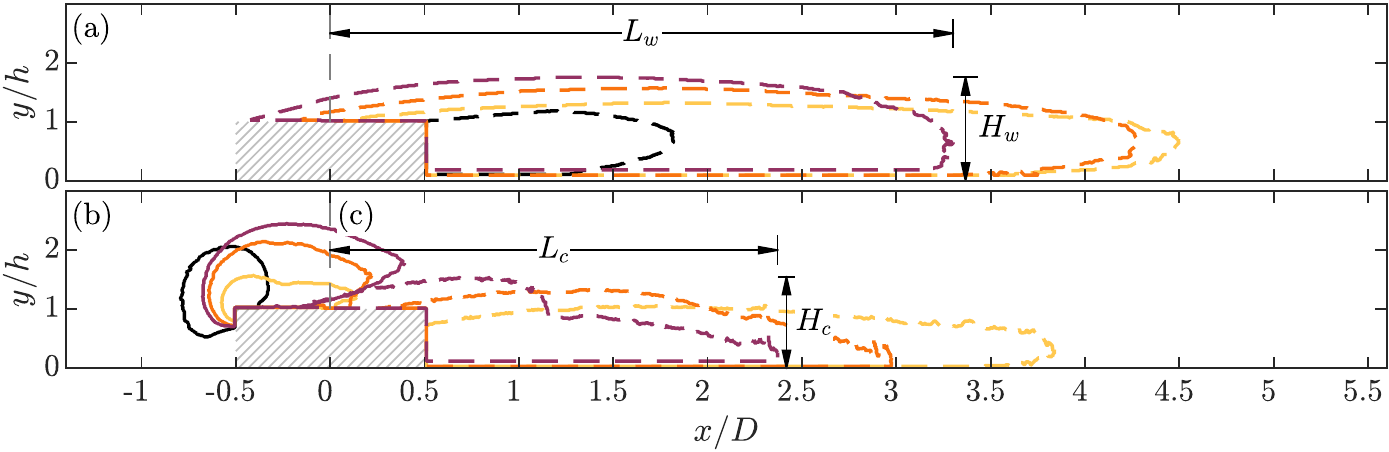}
	\caption{Contours of time-averaged (a) defect velocity $\Updelta U/U_{h/2} = 0.3$, (b) wall-normal velocity $V/U_{h/2} = 0.1$ at $x/D \leq 0.5$, and (c) scalar concentration $C/C_s = 3 \times 10^{-4}$. Colour scheme for each case (C39, C64, C133, and CS) is described in table~\ref{tab:cases}. The horizontal and vertical extent of the wake are defined by $L_w$ and $H_w$, respectively, while $L_c$ and $H_c$ are the horizontal and vertical extent of the scalar dispersion. Hatched regions illustrate the location of the patch.}
	\label{fig:contours}

	\includegraphics[width=14.4cm, height=10cm, keepaspectratio]{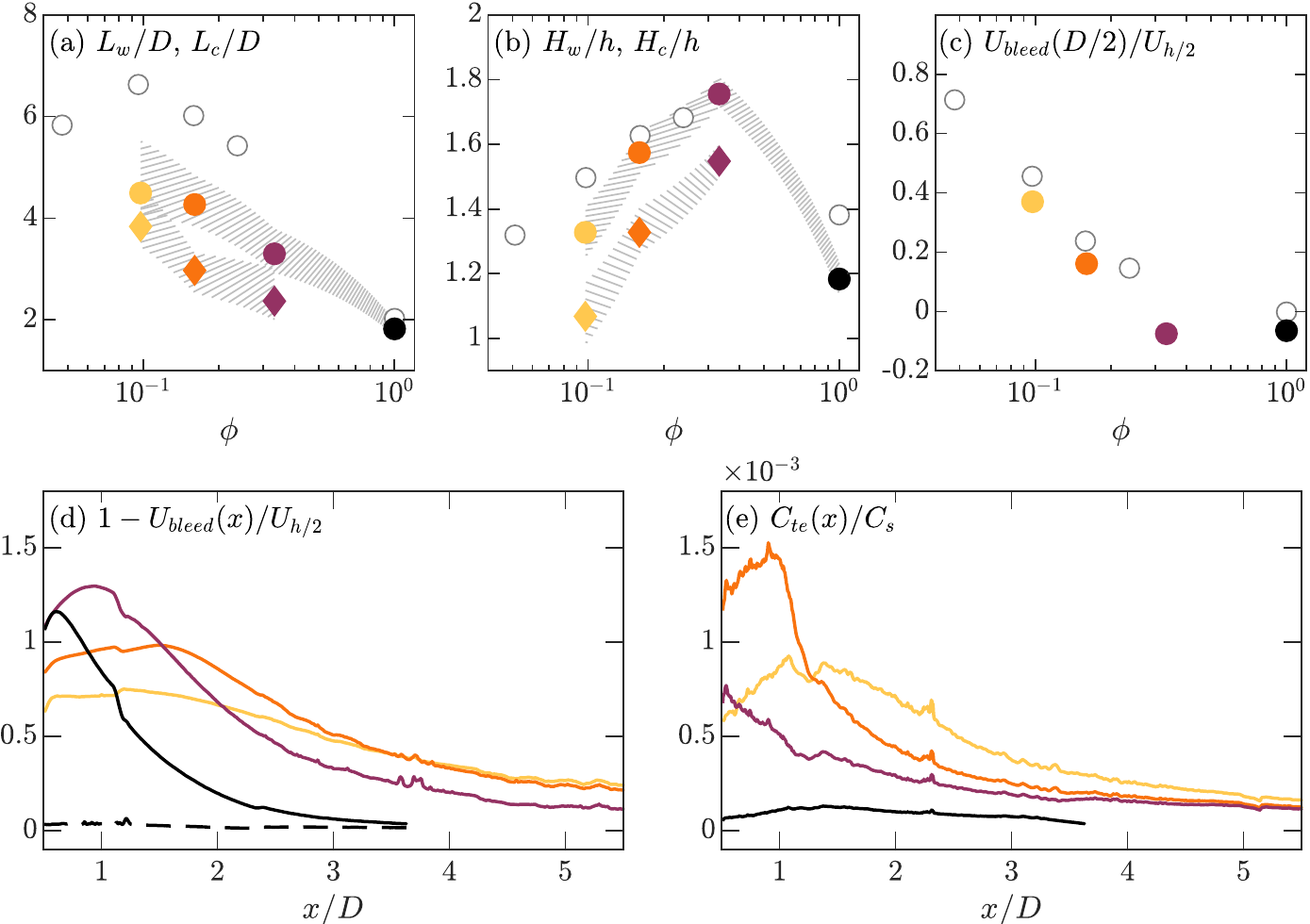}
	\caption{(a) Normalised length of wake $L_w/D$ ($\bullet$) and scalar dispersion $L_c/D$ ($\blacklozenge$) (as illustrated in figure~\ref{fig:contours}) of all test cases as a function of solidity $\phi$, (b) height of wake $H_w/D$ ($\bullet$) and scalar dispersion $H_c/D$ ($\blacklozenge$), and (c) integrated velocity bleeding $U_{bleed}/U_{h/2}$ at the trailing edge $x = D/2$. In (a,b), hatched regions illustrate the sensitivity to the threshold level defining the length/height ($\pm 20\%$). In (c), \textcolor{black}{$\circ$}: same quantities measured in \cite{nicolai2020} at $Re_{\theta} \approx 60000$. (d) $1-(U_{bleed}/U_{h/2})$ as a function of $x$ downstream of the patch, \dashedline: the reference smooth-wall case TBL. (e) Integrated scalar concentration $C_{te}/C_s$ as a function of $x$. Colour scheme for each case (C39, C64, C133, and CS) is described in table~\ref{tab:cases}.}
	\label{fig:te_bleed}
\end{figure}

Figures~\ref{fig:U} and~\ref{fig:C} show the dependency of wake and scalar concentration downstream of the porous bodies as a result of TE bleeding. We attempt to measure the extent of the wake by utilising a similar method in \cite{nicolai2020}. First, we define velocity defect as $\Updelta U = U_{bl} - U$, where $U_{bl}(y)$ is the incoming boundary layer velocity profile upstream of the patch (at $x/D = -1.3$), and arbitrarily choose a threshold for the magnitude of $\Updelta U$ as a wake definition. Figure~\ref{fig:contours}(a) shows the contours of $\Updelta U/U_{h/2}$ for all porous cases and the solid case (table~\ref{tab:cases}) at the chosen threshold of 0.3. With this threshold, we are able to define the length and height of the wake, $L_w$ and $H_w$, respectively (figure~\ref{fig:contours}a). $L_w$ is defined as the length of wake measured from the centreline of the patch ($x = 0$) to the $x$-maxima of the chosen contour level, while $H_w$ is the height of wake measured from the floor of the tunnel ($y = 0$) to the $y$-maxima of the same contour level. Figures~\ref{fig:te_bleed}(a,b) show the normalised length and height of the wake, $L_w/D$ and $H_w/h$, respectively, as a function of solidity $\phi$. The figures show that $L_w$ decreases with increasing $\phi$, but $H_w$ increases (and possibly reaches a maxima at certain value of $\phi$) and then decreases as $\phi \rightarrow 1$, where the obstacle is no longer porous and there is no flow escaping from the top of the patch. The wake `blow-up' (i.e. taller wake) in higher $\phi$ has also been recorded in \cite{taddei2016,nicolai2020}. These tendencies of $L_w$ and $H_w$ are insensitive to the change in threshold, as illustrated by the hatched regions in figures~\ref{fig:te_bleed}(a,b), which correspond to the same quantities measured when the threshold for $\Updelta U/U_{h/2}$ is varied by $0.3 \pm 20\%$. As a comparison, $L_w$ and $H_w$ measured in \cite{nicolai2020} are shown in white-filled circles in figures~\ref{fig:te_bleed}(a,b). The study involved similar arrays of cylindrical patches of various $\phi$ ($0.018 \leq \phi \leq 1$) that impinged a developing turbulent boundary layer at $Re_{\theta} \approx 60000$ (20 times higher than that of the current study). It should be noted that the measured $L_w$ and $H_w$ are greater in magnitude as the wake is defined at a lower threshold of $\Updelta U/U_{h/2} = 0.2$ instead of 0.3 due to the increasing streamwise length of the FOV ($8D$ instead of $6.89D$). It should also be noted that the values of $H_w$ (white-filled circles in figure~\ref{fig:te_bleed}b) are scaled by a factor of $3.58 h/\delta_{bl}$, where 3.58 is the ratio of incoming boundary layer thickness to cylinder height $\delta_{bl}/h$ in \cite{nicolai2020}. However, the differences between studies do not change the observed trend in $L_w$ and $H_w$ as a function of $\phi$.      

Similarly, a threshold can also be applied to the scalar concentration field $C$ to measure the extent of scalar dispersion downstream of the patch. Figure~\ref{fig:contours}(b) shows the contours of $C/C_s$ for all porous cases and the solid case (table~\ref{tab:cases}) at the chosen threshold of $3 \times 10^{-4}$. The length and height of scalar dispersion $L_c$ and $H_c$, respectively, are measured from the centreline of the patch $x = 0$ to the $x$-maxima of the chosen contour level and from the the floor of the tunnel $y = 0$ to the $y$-maxima, respectively. Filled diamonds in figures~\ref{fig:te_bleed}(a,b) show the measured $L_c$ and $H_c$ as a function of $\phi$. Similar to those of the wake extent, $L_c$ and $H_c$ decreases and increases, respectively, with increasing $\phi$. It should be noted that although scalar concentration data are available for case CS, the magnitude of $C$ downstream of the patch is lower than that of the chosen threshold (see figure~\ref{fig:C}d) and thus we are unable to measure $L_c$ and $H_c$ of this case. 

We further quantify the velocity of TE bleeding $U_{bleed}$. $U_{bleed}$ is defined as $U$ downstream of the patch ($x \geq D/2$) integrated over $y$. This definition can be extended further to the scalar concentration at the trailing edge $C_{te}$, such that
\begin{equation}
	\left .
	\begin{array}{r l}
		U_{bleed}(x) &= \frac{1}{h} \int_{0}^{h} U(x,y) \; \mathrm{d}y \quad \\
		C_{te}(x) &= \frac{1}{h} \int_{0}^{h} C(x,y) \; \mathrm{d}y \quad
	\end{array}
	\right\} x \geq D/2
\end{equation}
Figure~\ref{fig:te_bleed}(c) shows the magnitude of $U_{bleed}$ at the trailing edge, $x = D/2$, as a function of $\phi$. White-filled circles are $U_{bleed}$ measured by \cite{nicolai2020} for similar cylindrical arrangements at $Re_{\theta} \approx 60000$. Despite the difference in the order of $Re$, both \cite{nicolai2020} and the present study show that $U_{bleed}$ decreases with $\phi$ at the trailing edge. Figures~\ref{fig:te_bleed}(d) and~\ref{fig:te_bleed}(e) further show the defect $1-U_{bleed}/U_{h/2}$ and $C_{te}/C_s$ as functions of $x$ downstream of the patch ($x \geq D/2$). Both figures show the decay of wake and scalar concentration with $x$, with (generally) faster decay as $\phi$ increases in the porous cases. In particular, $U_{bleed}$ in figure~\ref{fig:te_bleed}(d) shows that the solid case CS recovers the fastest to the upstream condition (i.e., the incoming flow, dashed line is the smooth-wall case TBL), followed by the highest solidity case C133. We note that the decay of the wake for the porous and solid cases (figure~\ref{fig:te_bleed}d) and the decay of $C$ for the porous cases (figure~\ref{fig:te_bleed}e) seem to follow a certain power law, possibly similar to that of \cite{wygnanski1986} for various solid and porous wake generators, which is left for future works. The exception to this is the scalar concentration downstream of case CS (black solid line in figure~\ref{fig:te_bleed}e), which is approximately constant in $x$ and much lower in magnitude compared to the porous cases.    

\begin{figure}
	\centering\includegraphics[width=14.4cm, height=5cm, keepaspectratio]{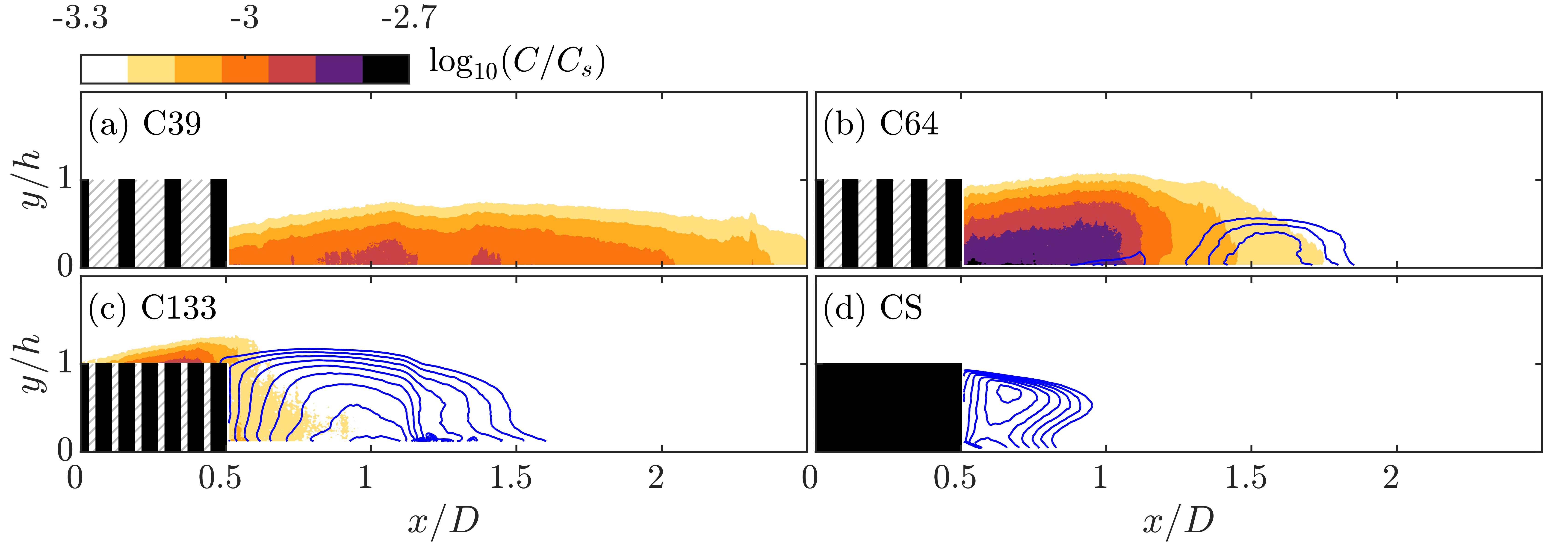}
	\caption{Contours of scalar concentration $\log_{10}(C/C_s)$ for cases: (a) C39, (b) C64, (c) C133, and (d) CS. Blue solid lines (\solidline[blue]) are negative streamwise velocity, in (b): $U/U_{h/2} = -0.06, -0.04, -0.02$ and in (c,d): $-0.4, -0.36, ... -0.04$.}
	\label{fig:recirc}
\end{figure}   

A closer look into figure~\ref{fig:te_bleed}(e) reveals a curious observation: the integrated scalar concentration of case C64 peaks at $x/D \approx 1$, and the magnitude of this peak is much higher than that of cases with lower and higher $\phi$ (C39 and C133). Figures~\ref{fig:recirc}(a--d) show the coloured contours of $C$ at the trailing edge for all porous and solid cases, with blue contour lines superimposed in the figures showing $U < 0$, which corresponds to flow recirculation. For case C64, figure~\ref{fig:recirc}(b) shows a high scalar concentration (purple-shaded region) at the trailing edge and up to $x/D \approx 1$, which is responsible for the spike in figure~\ref{fig:te_bleed}(e), immediately followed by the emergence of a recirculation bubble. In other words, the scalar is trapped between the porous patch upstream and the recirculation bubble that prevents the scalar from convecting downstream. By contrast, case C39 (figure~\ref{fig:recirc}a) that has lower solidity shows no recirculation bubble, and thus the scalar is freely transported downstream of the patch. Case C133 (figure~\ref{fig:recirc}c), on other hand, shows the recirculation bubble being shifted to the trailing edge. The scalar in this case is largely prevented from escaping towards the trailing edge (this explains the decrease in $L_w$ and $L_c$ as solidity increases, see figure~\ref{fig:te_bleed}a). Conservation laws dictate two possible outlets left for the flow: lateral and vertical bleeding. The higher magnitude of $C$ above the patch, upstream of the bubble ($y/h \geq 1$, $0 \leq x/D \leq 0.5$) confirms increasing vertical bleeding in higher solidity, while the increase of lateral bleeding with $\phi$ has been confirmed in \cite{zhou2019}. In some sense, the flow lingers downstream of the porous obstruction (characterised by high magnitude of $C$ in this region in case C64), it is as if the obstruction is extended in the streamwise direction. The length of this extension seems to be inversely proportional to $\phi$, as it is longer than the FOV for case C39 (or perhaps infinitely long), approximately equal to $1.25D$ for case C64 (figure~\ref{fig:recirc}b), and 0 for case C133 (figure~\ref{fig:recirc}c). Possibly, the extension length is akin to the length of the steady wake region observed in 2-D patches \citep{ball1996,zong2012,chen2012}. When the obstruction is impermeable (figure~\ref{fig:recirc}d), the mechanism differs. It acts as a solid body, shedding a pair of shear layers at the trailing edge (see contours of positive and negative $\Omega_y$ and $V$ at $0.5 \leq x/D \leq 1.5$ in figures~\ref{fig:U}d and~\ref{fig:C}d, respectively). 


\subsection{Vertical bleeding and the extent of scalar concentration}
\label{sub:v_bleed}

Contours of $V > 0$ above the patches are shown in solid lines in figure~\ref{fig:C}. The contours reveal a very different pattern of flow above the patches for porous cases (figure~\ref{fig:C}a--c) compared to the solid case CS (figure~\ref{fig:C}d). In case CS, flow separation is characterised by high magnitude of $V$ observed in the leading edge of the patch ($x/D = -0.5$). Compared to the solid case, the magnitude of $V$ in the leading edge is lower for all porous cases (figure~\ref{fig:C}a--c), as the flow is able to penetrate the patches instead of being diverted upward (see \citealp{taddei2016} for similar 3-D patches and \citealp{zhou2019} for suspended patches, note that the sign of $V$ is inverted in the latter). What is only observed in the porous patches, however, is that the presence of $V > 0$ region across the top surface of the patches ($-0.5 \leq x/D \leq 0.5$  and $y/h \geq 1$) due to vertical bleeding. The magnitude and extent of this region appear to be dependent upon solidity. As as comparison, the contours of $V/U_{h/2}$ for the porous cases in figure~\ref{fig:contours}(b) show that for the same magnitude ($V/U_{h/2} = 0.1$), $V$ is extended upward (in $y$) as $\phi$ increases.  

\begin{figure}
	\centering
	\includegraphics[width=14.4cm, height=5cm, keepaspectratio]{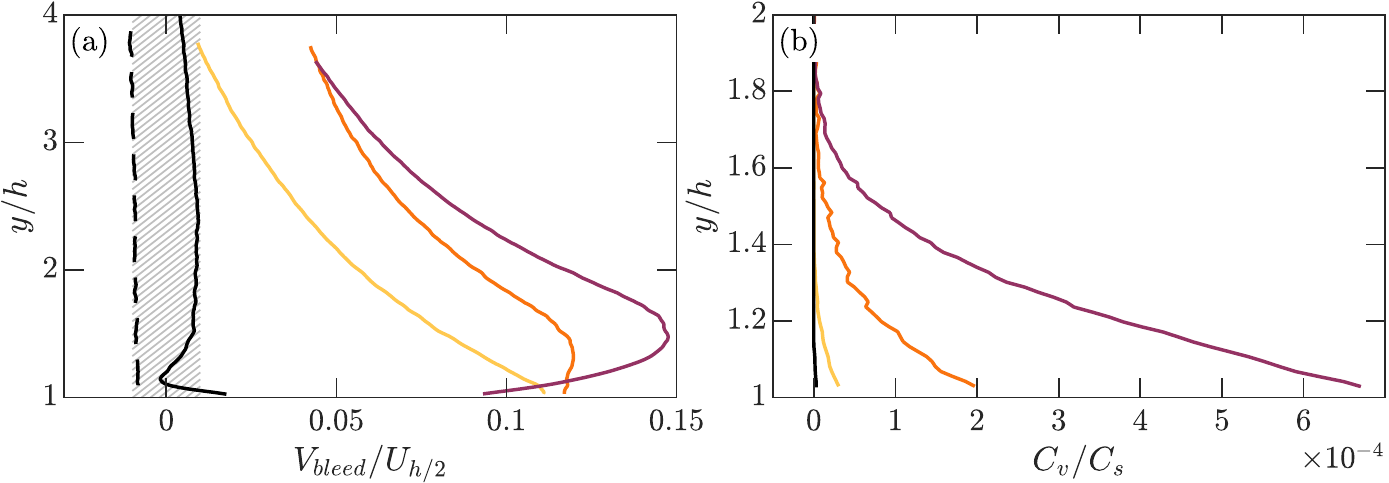}
	\caption{(a) Integrated vertical bleeding $V_{bleed}/U_{h/2}$ above the patch ($y/h > 1$) as a function of $y$. \dashedline: the reference smooth-wall case TBL. Hatched region shows $V_{bleed} \pm 1 \%$ of $U_{h/2}$. (b) Integrated scalar concentration above the patch $C_v/C_s$. Colour scheme for each case (C39, C64, C133, and CS) is described in table~\ref{tab:cases}.}
	\label{fig:v_bleed}
\end{figure}

Similar to the analysis in \S\ref{sub:te_bleed}, we quantify the magnitude of vertical bleeding as the integrated $V$ above the patch ($-D/2 \leq x \leq D/2$ and $y \geq h$), and this can be further extended to include the scalar concentration,
\begin{equation}
	\left .
	\begin{array}{r l}
		V_{bleed}(y) &= \frac{1}{D} \int_{-D/2}^{D/2} U(x,y) \; \mathrm{d}x \quad \\
		C_{v}(y) &= \frac{1}{D} \int_{-D/2}^{D/2} C(x,y) \; \mathrm{d}x \quad
	\end{array}
	\right\} y \geq h
\end{equation}  
Figures~\ref{fig:v_bleed}(a,b) show the integrated vertical bleeding $V_{bleed}$ and scalar concentration $C_v$ as functions of $y$ above the patch. Both figures show higher magnitude of $V_{bleed}$ and $C_v$ closer to the patch, which decay further towards the freestream. Dependency on $\phi$ is apparent on both. $V_{bleed}$ in figure~\ref{fig:v_bleed}(a) generally increases with $\phi$, with a significantly smaller magnitude (within $\pm 1\%$ of the incoming flow $U_{h/2}$, see hatched region in figure~\ref{fig:v_bleed}a) for CS and TBL (black solid and dashed lines in figure~\ref{fig:v_bleed}a, respectively).  For case CS, small $V_{bleed}$ is the result of averaging between high magnitude of $V > 0$ in the leading edge and $V < 0$ above the patch (see contour lines in figure~\ref{fig:C}d), while for case TBL ($V = 0$) this can be attributed to the error of the measurements. The integrated scalar concentration in figure~\ref{fig:v_bleed}(b) is 0 for the solid case CS (since there is no vertical bleeding, see black solid line in figure~\ref{fig:v_bleed}b), while for the porous cases it increases with $\phi$ and decays to 0 further from the patch.      

\begin{figure}
	\centering\includegraphics[width=14.4cm, height=3.5cm, keepaspectratio]{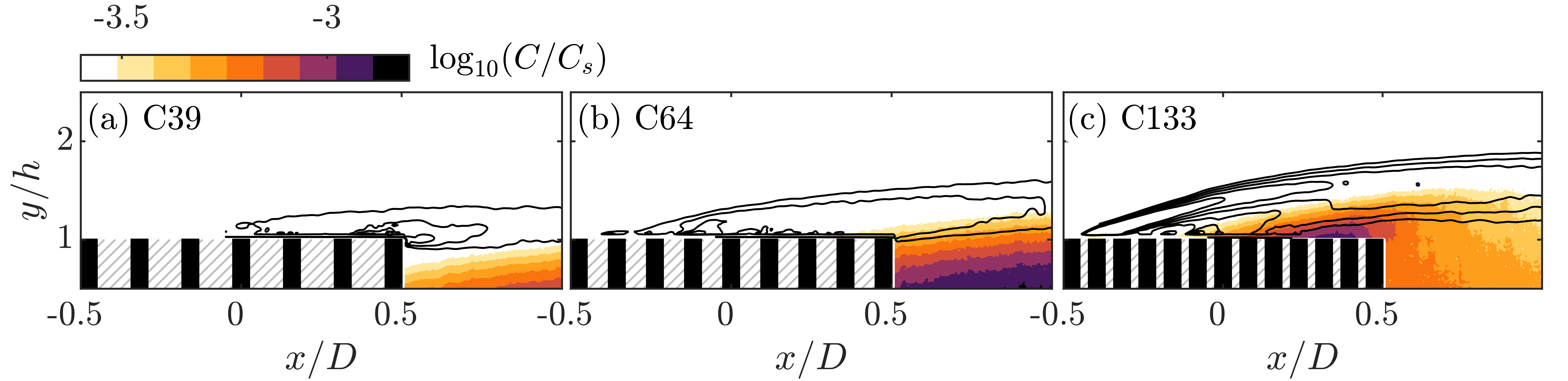}
	\caption{Contours of scalar concentration $\log_{10}(C/C_s)$ for cases: (a) C39, (b) C64, (c) C133, and (d) CS. Black solid lines (\solidline) are the normalised vorticity in $y$ axis $\Omega_y D/U_{h/2} = -8, -7, ..., -4$.}
	\label{fig:shear}
\end{figure} 

In addition to increasing vertical bleeding with $\phi$, we also observed that the maxima of $C_v$ in figure~\ref{fig:v_bleed}(b) occurs at the top surface of the patch $y/h  = 1$ for all porous cases, but the maxima of $V_{bleed}$ in figure~\ref{fig:v_bleed}(a) occurs further from the patch (for example, it occurs at $y/h \approx 1.5$ for case C133 in figure~\ref{fig:v_bleed}a). Figures~\ref{fig:shear}(a--c) offer a closer look of the flow above the porous patches. The coloured contours are the scalar concentration $C/C_s$ in logarithmic scale and the solid black lines are the contours of $\Omega_y < 0$, representing the shear layer forming above the patches. In case C39 (figure~\ref{fig:shear}a) the shear layer shields a region of (relatively) high scalar concentration (orange-shaded contours in figure~\ref{fig:shear}a) below the layer. As solidity increases in case C64 (figure~\ref{fig:shear}b), higher $C$ is observed downstream of the patch (purple-shaded coloured contours), which coincides with elevation of the shear layer compared to that of C39 in figure~\ref{fig:shear}(a). In the case with the highest solidity (C133 in figure~\ref{fig:shear}c), recirculation occurs on the trailing edge (figure~\ref{fig:recirc}c) followed by higher $C$ above the patch compared to that downstream of the patch (figure~\ref{fig:shear}c). Here the shear layer is elevated further upward, shielding the observed high $C$ above the patch ($0.25 \lesssim x/D \lesssim 0.5$). Similar to the analogy of `artificial' streamwise extension in the porous obstructions discussed in \S\ref{sub:te_bleed}, here there seems to be an artificial \emph{upward} extension of the obstruction as $\phi$ increases. This perhaps explains the phenomenon observed in figure~\ref{fig:v_bleed}, that the maxima of vertical bleeding occurs further from the patches but the maxima of scalar concentration occurs exactly at the top surface of the patches.     

\vspace*{-5pt}
\section{Conclusions and Recommendations}

We conduct planar PIV-PLIF measurements of the flow surrounding 3-D, circular porous patches of various solidity $\phi$ ($0.098 \leq \phi \leq 1$) to study the characteristics of flow in the wake of and above the patches. Porosity is generated by mounting rigid cylinders of height $h$ and diameter $d$ on the patch (whose diameter is $D$) and the variation in $\phi$ is achieved by systematically increasing the number of cylinders in a patch. The patches are fully submerged within the incoming flow, with a fluorescent dye injected from the floor of the tunnel, creating a point source upstream of the patches. Simultaneous PIV-PLIF measurements in $x$--$y$ plane provide streamise and wall-normal velocity components, as well as the concentration of the fluorescent tracer in the FOV.

Examination of the time-averaged velocity and scalar fields suggests that the trailing edge (longitudinal) and vertical flow bleeding from the obstructions decreases and increases with increasing $\phi$, respectively, as previously reported in \cite{taddei2016,zhou2019,nicolai2020}. In addition to that, we observe that $\phi$ affects the scalar concentration in the wake and above the patch. In particular, present results suggest that $\phi$ determines 
\begin{enumerate}[(i)]
	\item the horizontal extent of the wake ($L_w$) and scalar concentration ($L_c$) downstream of the porous patches, with $L_w$ and $L_c$ decreasing with $\phi$ (figure~\ref{fig:te_bleed}a). The decrease is attributed to the shift of recirculation bubbles towards the trailing edge of the patches as $\phi$ increases. High scalar concentration is observed downstream of the patch as the flow is trapped between patches and recirculation bubble, acting as an elongation of the obstructions in streamwise direction.
	\item the vertical extent of the wake ($H_w$) and scalar concentration ($H_c$) downstream of the porous patches, with $H_w$ and $H_c$ increasing with $\phi$ (figure~\ref{fig:te_bleed}b). As $\phi$ increases further, recirculation bubble occurs at the trailing edge of the patch and the flow is trapped in the interior of the patch. The flow largely escapes by means of vertical (and lateral) bleeding. High concentration of scalar, elevation of shear layer and $V$ above the patch suggest that the flow is trapped between the top surface of the patch, from which the flow bleeds, and the shear layer. Here we observe an artificial extension of the obstacle further upward (i.e. the patch is `taller' from the perspective of the flow) as $\phi$ increases.   
\end{enumerate}  

As a final remark, the mean scalar/flow fields  understanding both momentum \emph{and} scalar transport of the flow surrounding 3-D porous obstructions and building useful models for engineering design, prediction for meteorological or oceanographic flows, etc. The results presented in this study are limited to the time-averaged velocity and scalar components. However, we note that intermittency (i.e. discrepancies between the time-averaged and instantaneous flow fields) has been reported in \cite{cassiani2008} and \cite{taddei2016}. In the future, analysis about the fluctuating components ($u'$, $v'$, and $c'$) should be included to obtain the full picture of the flow characteristics. The decay of wake and scalar concentration in the trailing edge, shown in figures~\ref{fig:te_bleed}(d,e), indicate a possible wake model for both porous and solid cases, which should also be investigated in the future. 
	

\begin{Backmatter}
		

	\paragraph{Funding Statement} 
	The authors gratefully acknowledge the financial support from EPSRC (Grant Nos: EP/P021476/1 and EP/S013296/1).
			
	\paragraph{Declaration of Interests}
	The authors declare no conflict of interest.
		
	\paragraph{Author Contributions}
	CN designed, performed all measurements, and post-processed the data. DDW post-processed, analysed the data, and wrote the manuscript, with the supervision from BG. 
		
	\paragraph{Data Availability Statement}
	The data that support the findings of this study will be made available upon publication.	
		
	\paragraph{Ethical Standards}
	The research meets all ethical guidelines, including adherence to the legal requirements of the study country.
		
	\paragraph{Supplementary Material}
	Not available.
		
		
	\bibliographystyle{abbrvnat}
	\bibliography{flume_cylinder_paper_flow}
	 
		
\end{Backmatter}
	
\end{document}